\documentstyle[psfig,aps]{revtex}

\def \nl {\par \noindent}

\def\PLB{{\em Phys. Lett. B}}

\def\PR{{\em Phys. Rev.}}

\title{\bf Pion-Pion  Phase-Shifts and the Value of  Quark-Antiquark Condensate\\ in the Chiral Limit}
\author{Isabela P. Cavalcante\thanks{E-mail ipca@uerj.br} \ and
J. S\'a Borges\thanks{E-mail saborges@uerj.br}}
\address{
 Universidade do Estado do Rio de Janeiro\\ 
Rua S\~ao Francisco Xavier, 524, Maracan\~a, 
Rio de Janeiro, Brazil}

\begin{document}

\maketitle

\begin{abstract}

We use low energy pion-pion  
phase-shifts in order  to make distinction between the alternatives
for the value of  the quark-antiquark condensate $B_0$\ in the chiral
limit. We will consider the amplitude up to and including ${\cal
O}(p^4)$ contributions within the Standard and Generalized Chiral
Perturbation Theory  
frameworks. They are unitarized by means of Pad\'e approximants in order to  fit experimental phase-shifts in the resonance region.
As the best fits correspond to  $\alpha = \beta = 1$, we conclude that
pion-pion phase-shift analysis favors  the standard ChPT scenario,
which assumes just one, large leading order parameter $\langle \bar q q
\rangle_{_0}$.
 
\vspace{.5cm}

\nl PACS numbers: 12.39.Fe, 13.75.Lb. 
\end{abstract}

\section{Introduction}

Strong interaction phenomena at low energies are very constrained by 
chiral symmetry and by the structure of the Quantum Chromodynamics 
(QCD) ground state, which involves the quark and the gluon condensates. 
Massless QCD ($m_u = m_d = m_s = 0$) is  symmetric under the group
$SU(3)_L\times SU(3)_R$ that is  
spontaneously broken to $SU(3)_V$, what implies the existence of eight
Goldstone bosons coupled to the corresponding conserved axial quark
currents.  
The coupling strength   is related to the pseudo-scalar meson decay 
constant $F$, which in the chiral limit is of order $100$~MeV,  
much smaller than  typical hadron masses, say $1$~GeV. 

One may ask what is the size of  the quark condensate parameter 
$$B_0 = - \frac {\langle\,  \bar q q \, \rangle_{_0}} {F^{2}}.$$ 
Lattice calculations allow one to access this information,
which is needed, for instance, in the applications of QCD sum
rules. Another phenomenological approach that refers to this quantity
is Chiral Perturbation Theory (ChPT) \cite{Leu}\, the only theory for
low  energy QCD.  
ChPT  Lagrangian for meson processes is a series of terms of
different orders in the covariant derivatives of Goldstone fields and
their masses. However, the order in the external momenta $p$ of the
contribution of each term of the Lagrangian to a given process
depends on how big  $B_0$ is, as will be explained in the sequence.  
In the standard ChPT,  $B_0$\
is assumed 
to be as large as $\Lambda_H \sim 1$~GeV.
On the other hand, by considering the possibility of a low value for
the quark condensate, some authors \cite{Ster1} have developed a
program named Generalized Chiral Perturbation Theory (GChPT).  As a
new feature, the series of terms in the  
Lagrangian keeps $B_0$ as an expansion parameter of  ${\cal O}(p)$, 
as well as the quark masses, which are in turn accounted for as ${\cal
O}(p^2)$ in the standard case.    

Our motivation in the present exercise 
is that, as it is well known, the 
values for the pion-pion  S-wave scattering length within the two
approaches   differ by 30\%, being 
the one form the generalized approach closer to the experimental value
than in the standard case.
Here we try to look at another phenomenological
consequence of these different frameworks. 
The standard and 
generalized scenarios can be  analyzed by a global fit of pion-pion
phase-shifts. In this case, in order to access  the resonance region of
pion-pion scattering using ChPT, one may wish to use
Pad\'e approximants, as e.g. advocated in Ref. \cite{dob}, giving rise
to the so called inverse amplitude  method (IAM). 

In the next section we present the ChPT pion-pion scattering  amplitude and the
corresponding partial-waves. In section \ref{sec:fit} IAM is described
and the fits to experimental data 
are shown. The last section includes our conclusion and final remarks.

\section{Standard and Generalized ChPT}

In the standard ChPT \cite{Leu}\,   the
leading contribution to  low energy   
pion-pion scattering reproduces 
the soft-pion amplitude obtained by Weinberg \cite{1}, given by
\begin{equation}
A(s,t,u)= \frac 1 {F^2} \left(s - 2 \hat m B_0 \right) ,
\label{alea}
\end{equation}
where  $\hat m =  ( m_u + m_d ) / 2 $.
The leading order value of the pion mass  in the standard ChPT is
denoted by $M$, with  
$$ M^2 =2  \hat m B_0. $$  

On the other hand, in  GChPT one gets the same structure as (\ref{alea}), with $B_0$ replaced by
$$B = B_0 + 2 \, m_s Z^S_0 , $$
so that (\ref{alea}) reads
\begin{equation}
A(s,t,u)= \frac 1 {F^2} \left(s - M^2 - r  \frac {M^4}{B_0^2} Z^S_0 \right).
\label{aleas}
\end{equation}
The new low energy constant (LEC) introduced here is $Z^S_0$, 
which is  the coefficient of  $ \left[
\mbox{Tr} \left( U^\dagger \chi + \chi^\dagger U   \right) \right]^2 $
in the term of order $p^2$ of the GChPT Lagrangian, whereas in the
standard case this term would be  actually accounted for as  ${\cal O}(p^4)$.
The expression  (\ref{aleas}) also depends on the ratio $r = m_s /
\hat m$, where $m_s$ is the strange quark mass. 

These differences between the two approaches are more important 
when including loop corrections to the total amplitude. 
 In this case, some terms that are considered to be of orders $p^6$ 
and $p^8$ in the standard approach contribute to the one loop result. 
The general form of the amplitude up to one loop
 can be   written as \cite{Ster2}
\begin{eqnarray}
\lefteqn{A(s,t,u)= \frac {\beta} {F_\pi^2} \left(s - \frac 4 3 M_\pi^2
\right) + \alpha \frac {M_\pi^2} { 3 F_\pi^2} }  \label{ag}\\
&+& \frac  1 { F_\pi^4} \left(s - 2 M_\pi^2 \right)^2 \lambda_1
+  \frac  1 { F_\pi^4} \left[ \left(t - 2 M_\pi^2 \right)^2  
+ \left(u - 2 M_\pi^2 \right)^2 \right] \lambda_2 \nonumber \\
&+& \frac 1 {6 F_\pi^4}  \left\{ \left[ 4 \left[ \beta \left(s - \frac 4 3
M_\pi^2 \right) + \frac 5 6 \alpha 
M_\pi^2 \right]^2 - \left[ \beta \left(s - \frac 4 3 M_\pi^2 \right) -
\frac 2 3 \alpha  
M_\pi^2 \right]^2 \right] \bar J (s)  \right. \nonumber \\
& & \hspace{.3cm} + \left.  \left[ \frac 1 2 \left[ 
3 \left( \beta \left(t - \frac 4 3 M_\pi^2 \right) - \frac 2 3 \alpha 
M_\pi^2 \right)^2 + \beta^2 \left(s - u \right) 
\left( t - 4 M_\pi^2 \right) \right]  \bar J (t)   
+ \left( t \leftrightarrow u \right) \right]   \right\} , \nonumber 
\end{eqnarray}
\nl where $\bar J(x)$ is the loop integral, given by
$$
16 \pi^2 \bar J(x) = 2 + \sigma (x)\ln{\frac {\sigma (x) - 1}{\sigma
(x) + 1}}, \qquad
\sigma(x) = \sqrt{\frac {x - 4 M_\pi^2}{x}}. 
$$

This structure is equivalent to the expression in Eq. 2.1 
of Ref. \cite{b84},   obtained long ago by S\'a Borges  in the context
of the Unitarization Program of Current Algebra (UPCA), if one takes
$\alpha$ and $\beta$ equal to one. Let us mention 
that UPCA  polynomials  in
energy are  different from those in eq. (\ref{ag}) 
because their coefficients are collections of
subtraction constants inherent to the dispersion relation technique,
whereas in ChPT they come from combinations of LECs 
and chiral logarithms.  

The isospin defined amplitudes  $T_I$\, for $I\, =\,  0, 1\,\,
{\hbox{and}}\, \,  2$ are
\begin{eqnarray*}
T_0 (s,t) &=& 3 A(s,t,u)\, +\, A(t,s,u)\, + \, A(u,t,s) \, , \\
T_1 (s,t) &=&  A(t,s,u)\, - \, A(u,t,s) \, , \\
T_2 (s,t) &=&  A(t,s,u)\, + \, A(u,t,s)\, ,
\end{eqnarray*}
which are
expanded in partial-wave amplitudes, as
$$ T_I(s,t)\, = \, \sum_{\ell} (2 \ell + 1 ) \, t_{\ell\, I}(s) \, P_\ell \, (\cos
\theta), $$
where $P_\ell$ are the Legendre polynomials. In the following we omit
the label $\ell$, because we just deal with S-wave  ($I=0\, , 2$) and
P-wave  ($I= 1$).

Elastic unitarity implies that,  for\,  $ 16 m_\pi^2 \ge s \ge 4
m_\pi^2$\, , 
\begin{equation}
\mbox{Im} \, t_{I} (s) = \frac 1 {16 \pi} \sigma (s) |  t_{I} (s) |^2,
\label{unit}
\end{equation}
which can be solved yielding 
\begin{equation}
 t_{I} (s) = \frac {16 \pi}{ \sigma
(s)} e^{i\, \delta_I(s)}\, \sin \delta_I(s),
\label{delta}
\end{equation}
where $ \delta_I(s)$ are the real phase-shifts.

\section{Fitting pion-pion phase-shifts}
\label{sec:fit}

Our strategy is to disentangle between standard and generalized
versions of ChPT by fitting experimental phase-shifts.  Since ChPT
partial-wave 
amplitudes  respect eq. (\ref{unit}) only approximately they   must
be unitarized. In order to do this we follow the procedure of the inverse
amplitude method (IAM) \cite{dob}, that amounts to writing 
\begin{equation}
\tilde t_I (s) = \frac {t_I^{(2)}(s)}{ 1 -   t_I^{(4)}(s)\,
/t_I^{(2)}(s)}, \quad I=0,1\ {\hbox{and}}\ 2\, , 
\label{ttil}
\end{equation}
where $t^{(2)}$\,  and\, $t^{(4)}$ are partial-wave projections of
respectively ${\cal
O}( p^2)$\,  and\,  
 ${\cal O}( p^4)$ terms in eq. (\ref{ag}).  

This expression allows one to access the resonance region for
pion-pion scattering if one fixes the free parameters $\lambda_1$ \,
and \, $\lambda_2$\, by fitting the phase-shifts defined in
eq. (\ref{delta}) to experimental data. We recall
that these parameters are collections of LECs.   
In ChPT, the LECs are not related to the explicit symmetry breaking
parameter and have to be obtained phenomenologically. The traditional
way to do that is to extract them from $K_{\ell\, 4}$ decay and from 
D-wave scattering length,
for instance. Here the LECs corresponding to the parameters obtained 
from our
fits are not expected to be exactly the  ones found in literature, but
not very far from these, since the low energy behaviour of IAM
amplitudes are approximately the same as in the non-corrected ones. 

In the general amplitude given by (\ref{ag}), besides the constants
$\lambda_1$ and $\lambda_2$,  that will be used to fit the
phase-shifts, we have also the constants $\alpha$ and $\beta$, which  
can be written in terms of several  LECs, 
such as $B_0$ and $Z^S_0$ as well as other ones 
coming from higher orders and 
loop contributions. In the standard case, $\alpha$ and $\beta$ are
kept very close to one, while in GChPT they can change within a 
much wider range. The allowed values of these constants
\cite{Ster2} depend on the ratio $r=m_s / \hat m$ and are very close
to the corresponding
ones of the standard ChPT for large values of $r$, namely greater than
25. Particularly, for a reasonable value, say $r=10$, 
the allowed ranges are $ \ 2.2 < \alpha <
3.3 \ $ and $ \ 1.12 < \beta < 1.24 \ $ approximately. 
Regardless the value of $r$,  $\alpha$ can vary from 1 to 4
and $\beta$, up to 1.3.

For each fixed choice of $\alpha$ and $\beta$ within the ranges
mentioned right above, we perform the fit of P-wave phase-shift, which
determines the combination $\lambda_2 - \lambda_1$. This value is then
introduced in the expression of $I=0$ S-wave in order to
determine the remaining parameter by fitting experimental
data. $I=2$ amplitude does not depend on any new parameter. 
For each set of $\alpha$ and $\beta$,  we could evaluate the  quality
of the fit, by considering the related $\chi^2$.

\begin{figure}
\centerline{\psfig{figure=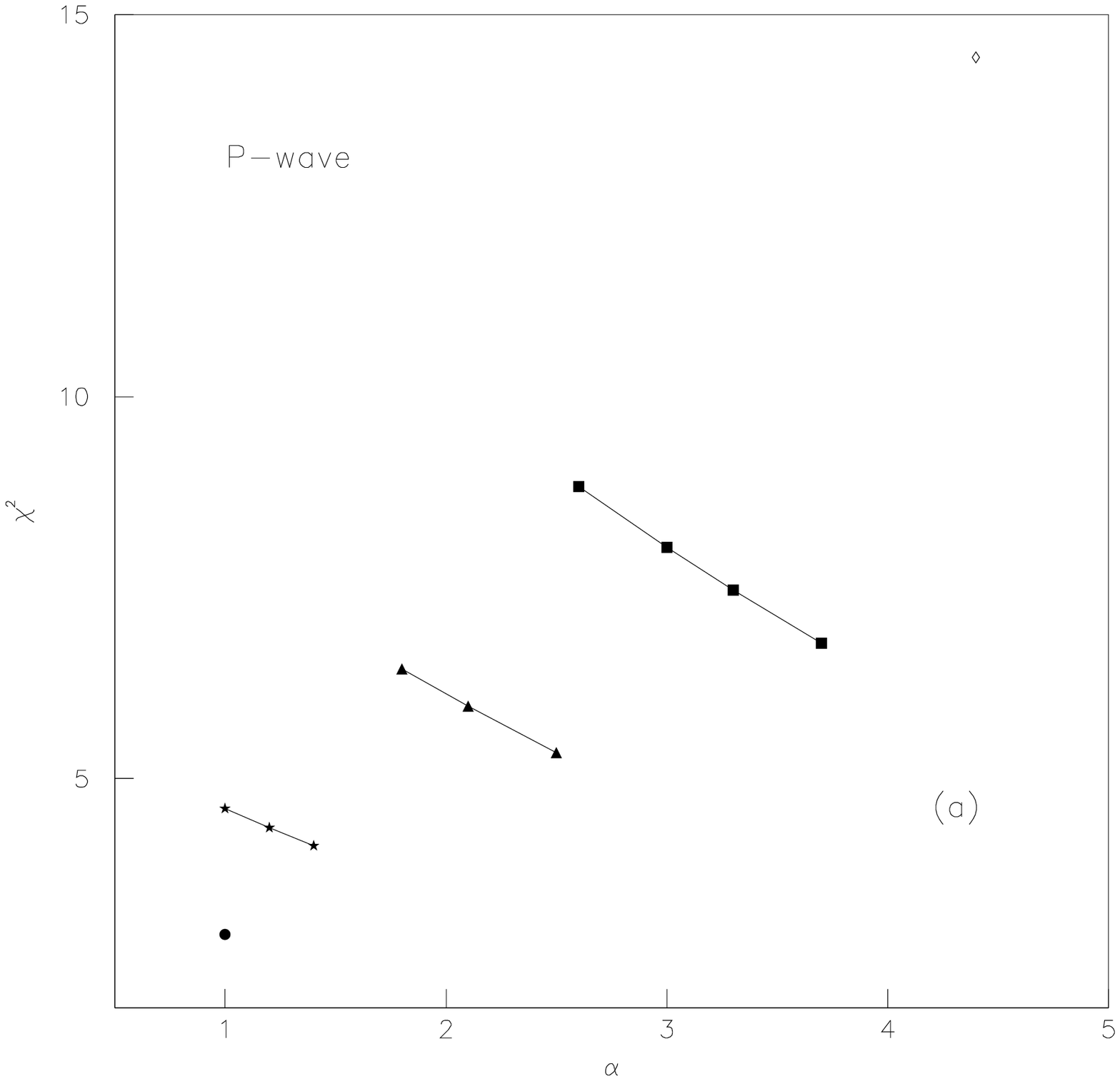,height=9cm}\hspace{.5cm} 
\psfig{figure=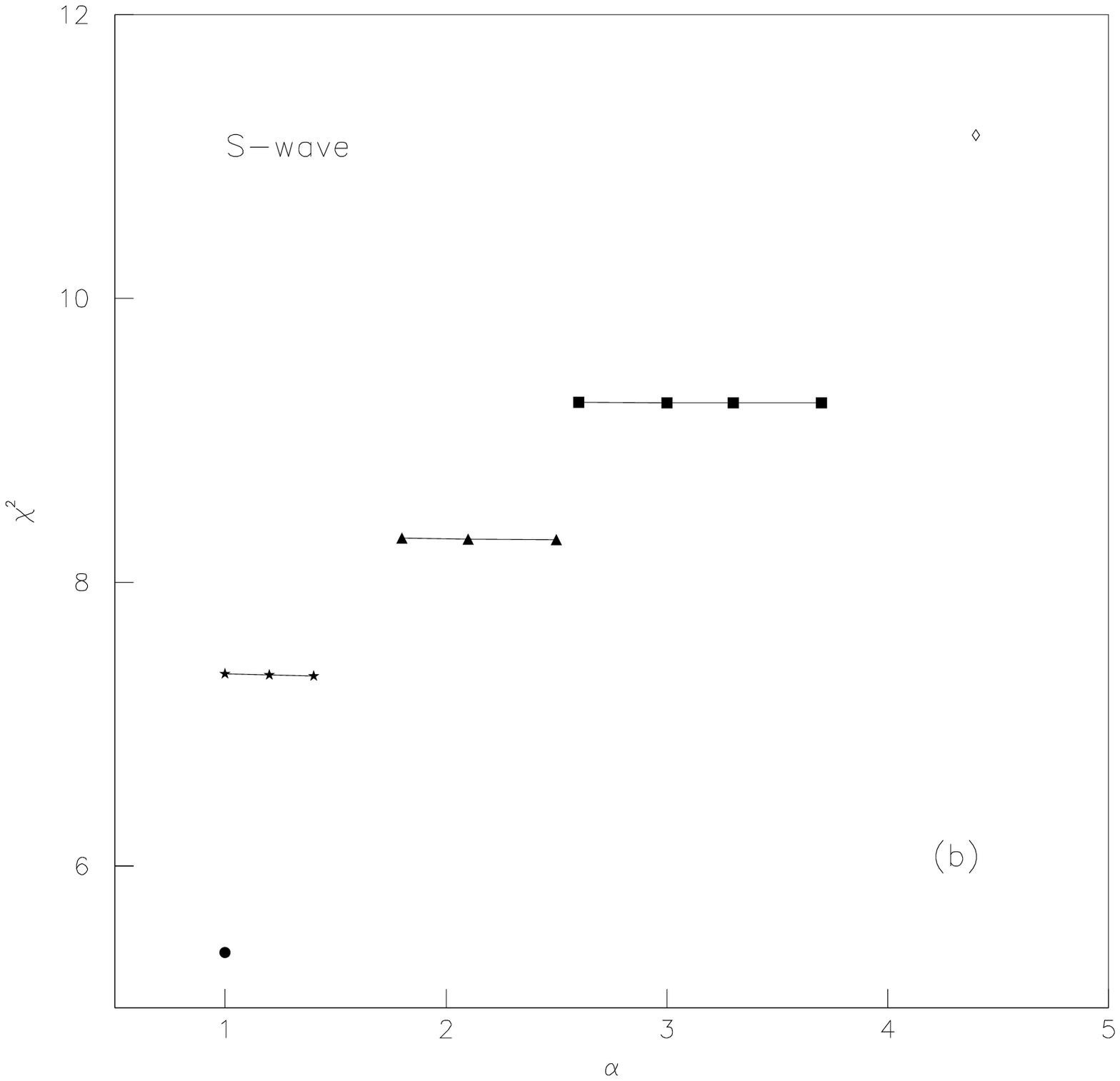,height=9cm}}
\caption{$\chi^2$ of (a) P-wave and (b) and S-wave fits to 
pion-pion phase-shifts,  as a function of $\alpha$
(adimensional); each symbol corresponds to a singular value of $\beta$. }
\label{chi2}
\end{figure}

\begin{figure}
\centerline{\psfig{figure=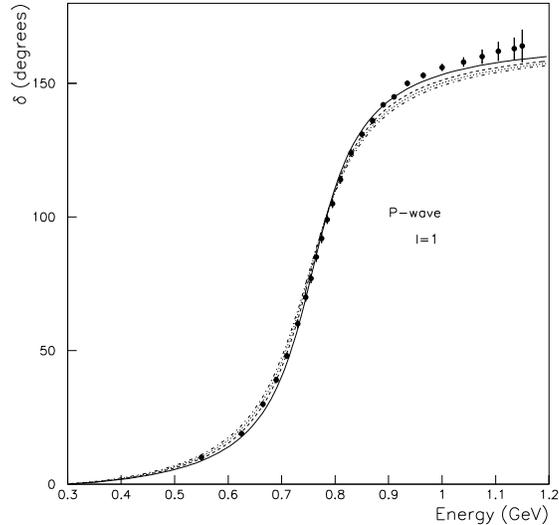,height=8cm}}
\caption{P-wave pion-pion phase-shift, in degrees, as a function of
center of mass energy, in GeV; experimental data (bullets) 
and results from fits, 
for $\alpha$ and $\beta$ as given in the Table.}
\label{pplot}
\end{figure}

\begin{figure}
\centerline{\psfig{figure=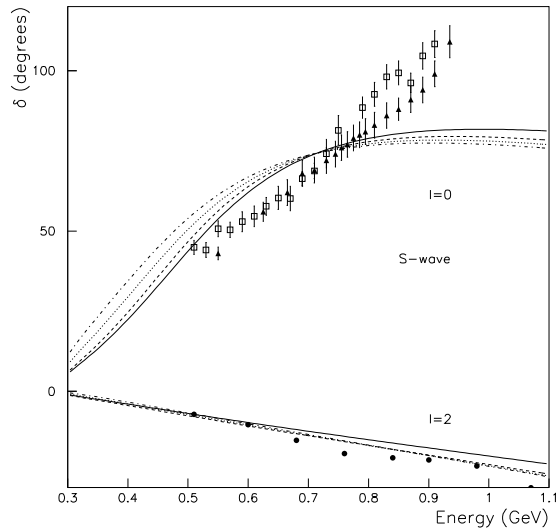,height=8cm}}
 \caption{S-wave pion-pion phase-shift, in degrees, as a function of
 center of mass energy, in GeV; experimental data for $I=0$ (squares and
 triangles) 
and $I=2$ (bullets) and results from fits of $I = 0$ wave, 
for $\alpha$ and $\beta$ as given in the Table.}
\label{splot}
\end{figure}

Fig. \ref{chi2} shows the $\chi^2$ of the fits for several choices of
$\alpha$  and $\beta$.  
One can see that, for a given value of   $\beta$, the $\chi^2\,$ is
not considerably modified within the range of  $\alpha$ values allowed
according to Ref. \cite{Ster2}. The
quality of fits is very dependent on $\beta$ and the best fit corresponds
to the bullet in the figure ($\alpha = 1$ and $\beta =1$).
According to this analysis, Figs. \ref{pplot} and \ref{splot} present
the resulting 
phase-shifts\footnote{Experimental data for P-wave were taken 
from Ref. \cite{pro}, for
$I=0$ S-wave from Ref. \cite{pro} (squares) and Ref. \cite{estra}
(triangles) and for $I=2$ S-wave from Ref. \cite{lost}.} of
respectively P- and S-waves, for only one
choice of $\alpha$ for each value of  $\beta$. 

The table presents the
values of parameters $\lambda_1 \, $  and  $\lambda_2 \, $ 
for P1- and S0-wave fits and the value of the related $\chi^2\,$'s.
Clearly the fits get poorer for large values of $\beta$. It  
 means that, for $r \simeq 10$  
the standard ChPT approach is more suitable than
GChPT to fit
$\pi \pi $ phase-shifts in a wide energy range. Nevertheless 
if one accepts a large enough value for $r$, then the two approaches
 are indistinguishable within  this method.

\section{Conclusion}

As a summary,  we have presented a study motivated by discrepancies
between the standard and generalized ChPT predictions for S-wave
pion-pion  scattering length.  Our concern was to verify whether
pion-pion phase-shifts could distinguish between the large and low
value of the quark condensate. In  order to achieve it, we have fitted
experimental data using the expressions following from the two
scenarios. Our work favors the
standard approach of ChPT, which assumes a large vacuum expectation 
value for the
quark-antiquark condensate operator.


\begin{table}
\begin{center} 
\begin{tabular}{l|l|c||l|l|l|l} 
 $\alpha$\ \ \ &$\beta$\ \ \  & line type &
$\lambda_1\, \times 10^3$ &  $\lambda_2 \, \times 10^2$ &  $\chi^2$ (P1)  & $\chi^2$ (S0)   \\
\hline
1.0 &   1.0  & solid &  -3.024    &1.1702  & 2.96  & 5.39 \\
1.0 &  	1.1 & dashed &  -3.268    &1.2971  &  4.61 & 7.36  \\  
2.6 &	1.2 & dotted &  -3.858    &1.4999  &  8.82  & 9.27  \\ 
4.4 &  	1.3 & dot-dashed & -4.562    &1.7235  &  14.4 & 11.2 
\end{tabular}
\end{center}
\caption{Fitted parameters $\lambda_1$ and $\lambda_2$ and
$\chi^2$ of the fits, corresponding to the various curves in Figs. 1
and 2.  The $\chi^2$ of the $N$-point fit is  
given, as usual, by the sum of deviations from data squared, weighted by
error bars squared, and divided by $(N-1)$.
}
\end{table}

\end{document}